\documentclass[aps,prl,reprint,superscriptaddress]{revtex4-1}

\usepackage{graphicx}

\usepackage{amsmath}

\usepackage{amssymb}

\begin{document}


\title{Frictional magneto-Coulomb drag in graphene double-layer heterostructure}


\author{Xiaomeng Liu}
\affiliation{Department of Physics, Harvard University, Cambridge, MA 02138, USA}

\author{Lei Wang}
\affiliation{Department of Mechanical Engineering, Columbia University, New York 10027, USA}

\author{Kin Chung Fong}
\affiliation{Raytheon BBN Technologies, Quantum Information Processing Group, Cambridge, MA 02138, USA}

\author{Yuanda Gao}
\affiliation{Department of Mechanical Engineering, Columbia University, New York 10027, USA}

\author{Patrick Maher}
\affiliation{Department of Physics, Columbia University, New York 10027, USA}

\author{Kenji Watanabe}
\affiliation{National Institute for Materials Science, 1-1 Namiki, Tsukuba 305-0044, Japan}

\author{Takashi Taniguchi}
\affiliation{National Institute for Materials Science, 1-1 Namiki, Tsukuba 305-0044, Japan}

\author{James Hone}
\affiliation{Department of Mechanical Engineering, Columbia University, New York 10027, USA}

\author{Cory Dean}
\affiliation{Department of Physics, Columbia University, New York 10027, USA}

\author{Philip Kim}
\affiliation{Department of Physics, Harvard University, Cambridge, MA 02138, USA}


\date{\today}

\begin{abstract}
Coulomb interaction between two closely spaced parallel layers of electron system can generate the frictional drag effect by interlayer Coulomb scattering. Employing graphene double layers separated by few layer hexagonal boron nitride (hBN), we investigate density tunable magneto- and Hall-drag under strong magnetic fields. The observed large magneto-drag and Hall-drag signals can be related with Laudau level (LL) filling status of the drive and drag layers. We find that the sign and magnitude of the magneto- and Hall-drag resistivity tensor can be quantitatively correlated to the variation of magneto-resistivity tensors in the drive and drag layers, confirming a theoretical formula for magneto-drag in the quantum Hall regime. The observed weak temperature dependence and $\sim B^2$ dependence of the magneto-drag are qualitatively explained by Coulomb scattering phase-space argument.
\end{abstract}


\maketitle


\begin{figure*}
 \centering
 \includegraphics[scale=0.55]{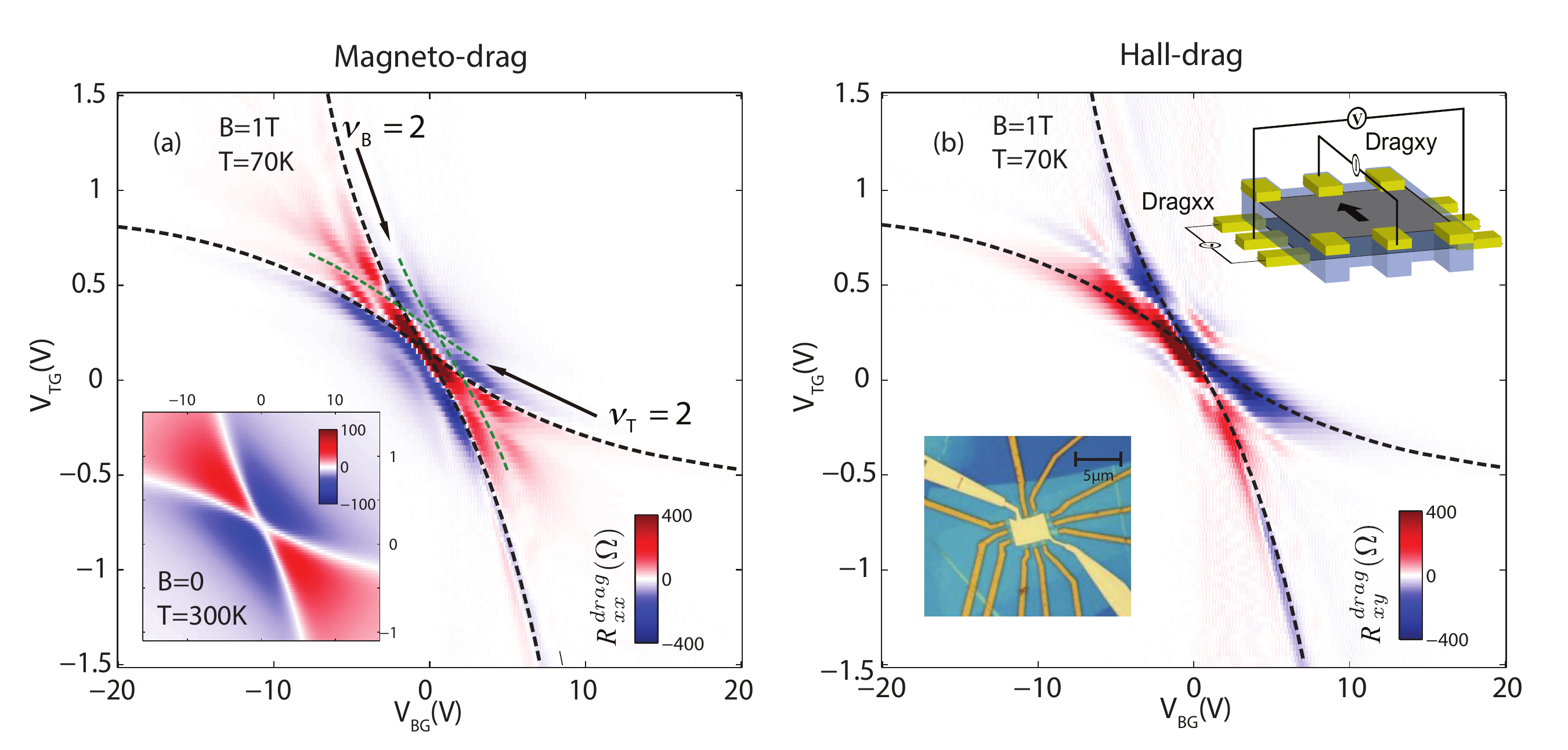}
 \caption{(color online)(a) Magneto-drag resistance as a function of top gate ($V_{TG}$) and bottom gate voltages($V_{BG}$) measured under a magnetic field of 1T and at a temperature of 70K. Black and green dashed lines mark charge-neutrality and $\nu=2$ of the individual layers, respectively. The lower inset shows the drag resistance at zero magnetic field and a higher temperature of 300K. (b) Hall-drag resistance as a function of gate voltages measured at $B=1$~T and $T=70$~K. Black dashed lines mark charge-neutrality of the individual layers as in (a). The upper insert shows the measurement schematics of the experiment. The lower inset shows an optical microscope image of the device used in this experiment. The scale bar corresponds to 5~$\mu$m.}{\label{F1}}
 \end{figure*}

Electronic double layers (EDL), consisting of two parallel conducting layers separated by a thin dielectric, provide a versatile platform to study interaction driven phenomenon in two-dimensional (2D) systems. For example, the Bose Einstein condensation of magneto-exciton in strongly interacting quantum Hall EDL has been discovered in GaAs EDL~\cite{Eisenstein2014,Eisenstein2004} and recently in graphene EDL~\cite{Liu2016, Li2016}. The EDL can also be used to study resonance tunneling ~\cite{Britnell2013}, proximity screening of disorder~\cite{Ponomarenko2011}, and penetration field ~\cite{Lee2014}.

Drag measurement in an EDL, i.e., applying current $I_{drive}$ in one of the layer (the 'active' drive layer) and probing induced voltage $V_{drag}$ in the other layer (the 'passive' drag layer), has been a useful to tool to characterize the interlayer Coulomb interaction. In a weakly coupled regime at a finite temperature $T$, the drag resistance $R_{drag}=V_{drag}/I_{drive}$ is typically dominated by momentum transfer through interlayer electron-electron (e-e) scattering. This frictional drag effect has been studied by both semiconductor~\cite{Gramila1991,Solomon1989,Sivan1992} and graphene EDLs~\cite{Gorbachev2012,Kim2011}. In general, under zero magnetic fields, the EDLs can be described by the Fermi liquid theory, and a semi-classical picture can explain the observed frictional drag effect~\cite{Narozhny2016}. In this regime two important features emerge: (1) drag is negative (positive) when two layers have the same (opposite) type of carriers, owing to the current and momentum relation; (2) drag resistance scales with temperature as $R_{drag} \propto T^2$, reflecting the increasing scattering phase space as temperature increases (Coulomb scattering phase-space argument) ~\cite{Narozhny2016, Gramila1991}.  Recent studies in graphene EDL suggest new drag mechanisms other than the above mentioned momentum drag also play important roles near the double charge neutrality point (CNP)~\cite{Gorbachev2012,Titov2013,Lee2016a,Li2016a,Song2013,Song2013a,Titov2013}.

In the presence of magnetic fields, the momentum transfer direction in the drag process is not aligned with the drive current direction, and thus drag voltages can be decomposed into magneto-drag (longitudinal component) and Hall-drag (transverse component). Moreover, under strong magnetic fields, quantized Landau levels (LLs) form in both layers, requiring consideration beyond the semiclassical description. Early experimental works in GaAs EDLs revealed that the sign of magneto-drag depends on the LL filling factor difference between the two layers~\cite{Feng1998,Lok2001}, which was not expected in a simple semiclassical model. Extending prior theoretical work based on the linear response theory~\cite{Kamenev1995,Flensberg1995,Narozhny2000,Narozhny2001}, von Oppen, Simon and Stern (OSS) proposed a theoretical approach to frictional drag under strong magnetic fields~\cite{VonOppen2001}. According to OSS, the drag resistivity tensor $\hat{\rho}^{drag}$ can be related to the density differential of the magneto-conductivity tensors $\hat{\sigma}$ in individual layers:
\begin{equation}
\hat{\rho}^{drag}\sim-\hat{\rho}^{p}\frac{d\hat{\sigma^{p}}}{dn^{p}}\frac{d\hat{\sigma^{a}}}{dn^{a}}
\hat{\rho}^{a}
\label{formula}
\end{equation}
Here, $\hat{\sigma}$ and $\hat{\rho}\equiv\begin{pmatrix}
\rho_{xx}&\rho_{xy}\\\rho_{yx} &\rho_{yy}
\end{pmatrix}$ are the magneto resistivity and conductivity tensors, respectively, $n$ is carrier density of each layer, and the superscripts $drag$, $a$, and $p$ stand for the drag, active (drive) and passive (drag) layers, respectively. The physical interpretation of this theory is that driving DC current on one layer creates asymmetry in the thermal density fluctuations in that layer. These density fluctuations are transferred to the drag layer through Coulomb interaction. Then the induced density fluctuations in the drag layer are rectified to a DC voltage. The differential conductivity enters the formula through rectification coefficient, which is proportional to $d\hat{\sigma}/dn$ assuming local current-voltage relation. This formula enables negative magneto-drag when the derivative of conductivity tensor meet the right condition. It also predicts that Hall-drag could have the same magnitude as the magneto-drag.

Graphene double-layer devices provide an excellent material platform to investigate the magneto-drag in the quantum limit, owing to a wide range of gate tunability of individual layers, large LL separation, and small inter-layer distance. In this letter, we present the experimental investigation of frictional magneto- and Hall-drag in high mobility graphene double layers. The observed drag can be quantitatively related to the modulation of the measured conductivity tensors in individual layers, confirming the OSS theory. Magnetic field and temperature dependence of the drag effect further reveals the nature of the Coulomb interaction in quantized LLs of the EDL systems.

The devices used in this experiment consist of two monolayer graphene separated by a thin hexagonal boron nitride (hBN) spacer $\sim$4~nm, encapsulated by two thicker BN layers ($\sim$20~nm). The thickness of interlayer hBN was chosen to provide a strong Coulomb interaction between the layers without direct tunneling of carriers between graphene layers~\cite{Gorbachev2012}.
The hetero-structure is make using dry transfer method~\cite{Wang2013} and edge contacts are fabricated on individual graphene layers~\cite{Liu2016,Li2016}. Two devices with the similar device geometry used in our experiments provide qualitatively similar results. The low temperature (1.5~K) mobility of the bottom layer is $\sim$50~m$^2$/Vs and the top layer shows a slightly lower mobility of $\sim$20~m$^2$/Vs. The high mobility we achieved in this device allows us to observe quantum Hall effect (QHE) at magnetic field $B$ as low as 0.2~T in both layers.

\begin{figure*}
 \centering
 \includegraphics[scale=0.7]{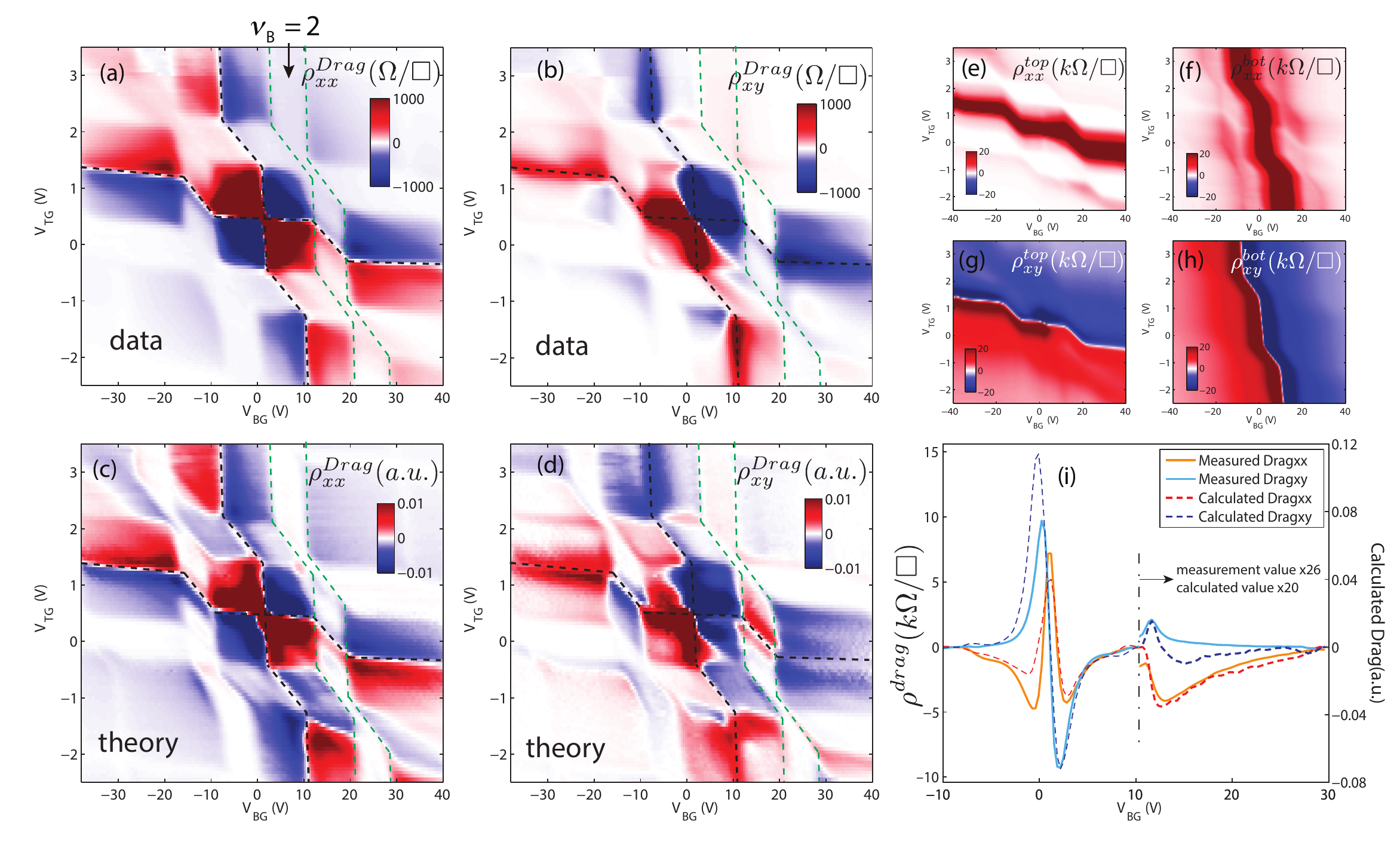}
 \caption{(a) and (b), measured magneto- and Hall-drag resistivity at T=70K, B=13T. (c) and (d), calculated magneto- and Hall-drag resistivity using Eq.~(1). The calculation is in arbitrary unit (a.u) due to the undetermined prefactor. The black dashed lines in (a-d) are charge neutrality lines of top and bottom layers. And the green outlined regions mark $\nu_{top}=2$ incompressible strip. (e) and (f), magneto-resistivity of top and bottom layers at T=70K, B=13T. (g) and (h), hall-resistivity of top and bottom layers under the same condition. (i) line-cut of measured and calculated magneto- and hall-drag along the equal-density line ($n_{T}=n_{B}$). The dashed vertical line around $V_{BG}\approx$10~V separates the first LL (N=1) region (right) from the zeroth LL (N=0) region (left). In the first LL region, measured drag is multiplied by a factor 26 for clarity while the calculation is multiplied by a factor 20 in order to make a good comparison to the measured values.} {\label{F2}}
 \end{figure*}

The drag measurements are performed by applying a small drive current $I_{drive}\sim$100~nA to the (active) drive layer and by measuring the drag voltages in the (passive) drag layer. Despite using a lock-in setup, the low frequency measurements (17.7Hz) essentially probe the DC drag response. To eliminate spurious signals originating from interlayer bias gating effect~\cite{Hill1996}, interlayer balancing is implemented in the drive layer (biasing current with bridge circuit)~\cite{Kellogg2005}. The Onsager reciprocity and linear response of the drag signal to $I_{drive}$ are confirmed in our experiment. Interlayer tunneling resistance is found to be larger than G$\Omega$ range. The magneto-drag resistance $R^{drag}_{xx}$ and Hall-drag resistance $R^{drag}_{xy}$ are obtained from the measured voltages across the passive (drag) layer divide by the drive current in the active (drive) layer, after (anti) symmetrizing ($R^{drag}_{xy}$) $R^{drag}_{xx}(B)$ with respect to magnetic field direction reversal. Voltages applied to the back gate ($V_{BG}$) and the top gate ($V_{TG}$) control the carrier density of the drive layer (top) $n_T$ and the drag (bottom) layer $n_B$.

Fig.1 shows $R^{drag}_{xx}$ and $R^{drag}_{xy}$ as a function of $V_{BG}$ and $V_{TG}$, measured at $T=$70~K and relatively low magnetic field $B=$1~T. The two black dashed lines crossing each other correspond to $\nu_T=0$ or $\nu_B=0$, the CNPs of each layer. The top (bottom) layer carrier density is mainly tuned by $V_{TG}$ ($V_{BG}$). These CNP lines also divide the $(V_{TG}$-$V_{BG})$ plane into four regions, e-e (top-right), h-h (bottom-left), e-h (bottom-right) and h-e (top-left). For magneto drag (Fig.1a), the sign of $R^{Drag}_{xx}$ follows the sign of drag at $B=0$ (Fig.~1(a) inset), i.e. the e-e and h-h regions show a negative drag signal, while the e-h and h-e regions exhibit a positive drag where $R^{Drag}_{xx}\approx$0 along the CNP lines dividing these regions. We also note that there is additional modulation in each regions, where some $R^{Drag}_{xx}\approx$0 lines running in parallel with CNP lines (examples are marked by green dashed lines). Further inspection in connection with the magneto-resistance measurements of each layer (which will be discussed later in detail) indicates that these lines are corresponding to $(V_{TG}, V_{BG})$ where either the active or passive layers are in the quantum Hall (QH) states with integer LL filling fraction $\nu_T$ or $\nu_B$. The vanishing $R^{Drag}_{xx}$ signal in these QH regions thus suggest the drag become inefficient as the bulk of either drive or drag layer becomes incompressible. The incompressible bulk results in zero density of state for interlayer Coulomb scattering. This observation is more pronounced at higher magnetic fields where stronger QHE appears with a wide range of incompressible regions in the $(V_{TG}$-$V_{BG})$ plane. Fig.2~(a) shows $\rho^{Drag}_{xx}(V_{TG},V_{BG})$ measured at B=13T, where the well-developed zig-zag shaped incompressible stripes of QH states can be identified with zero drag (for example the green dashed lines surround $\nu_{bot}=2$ incompressible strip where drag vanishes). The black dashed lines again mark the CNP of each layer. The zig-zag shape of the CNP and other incompressible stripes originate from difference of screening effect inside and outside of LLs (nearly perfect screening inside LLs) as we discussed above.

The corresponding Hall-drag $R^{drag}_{xy}$ measurements shows similar vanishing signals in the incompressible regions in the $(V_{TG}$-$V_{BG})$ plane as shown in Fig.~1(b). We note $R^{drag}_{xy}$ exhibits similar magnitude as $R^{drag}_{xx}$, confirming the prediction made by OSS in Eq.~(1). However, unlike $R^{drag}_{xx}$ whose sign is determined by the sign of carriers, $R^{drag}_{xy}$ undergoes sign changes within each quadrant. Contrary to $R^{drag}_{xx}$, $R^{drag}_{xy}$ does not vanish along the CNP lines. At higher magnetic field $B=13$~T (Fig.~2(b)), the incompressible QHE regions exhibit well-developed stripe regions of vanishing $\rho^{drag}_{xy}$ similar to $\rho^{drag}_{xx}$.


To compare density dependent magneto- and Hall-drag with Eq.(1), we need to obtain magneto-tensor $\hat{\rho}$ and $\hat{\sigma}$ as a function of density. Experimentally, we measured the longitudinal ($R_{xx}$) and transverse ($R_{xy}$) component of magneto-resistance on each layer and then converted them to $\hat{\rho}$ and $\hat{\sigma}$ using geometrical factors simulated by finite element method considering the device configuration. Fig.~2(e-h) are measured $\rho_{xx}$ and $\rho_{xy}$, the two independent components of $\hat{\rho}$, of the top and bottom layers as a function of the top and gate voltages $V_{TG}$ and $V_{BG}$. These data were taken at the same condition as the drag experiment shown in Fig.~2(a-b). Under strong magnetic fields, the relation between the density and $V_{T}$ and $V_{B}$ can be complicated due to the screening effect in LLs, resulting kinked striped incompressible regions. In general, the derivation of conductivity respect to density $\frac{d\hat{\sigma}}{dn}$ thus include derivation to both top and bottom gate:
\begin{equation}
\frac{d\hat{\sigma}}{dn}=\frac{d\hat{\sigma}}{dV_{BG}}\frac{dV_{BG}}{dn_{B}}+\frac{d\hat{\sigma}}{dV_{TG}}\frac{dV_{TG}}{dn_{T}}.
\end{equation}
However since $\frac{d\hat{\sigma}}{dn}$ is non-zero only in the compressible regions of ($V_{T}$, $V_{B}$) where the gating effects then decouple to each corresponding layer due to nearly perfect screening, resulting $n_{T}=C_{TG}V_{TG}/e$ and $n_{B}=C_{BG}V_{BG}/e$. Therefore, taking derivative respect to densities is same with respect to gate voltages times geometric capacitances.

Fig.~2(c-d) shows the computed drag resistivity $\rho_{xx}^{drag}$ and $\rho_{xy}^{drag}$ obtained from ${\hat\rho}^{drag}$ by applying experimental obtained ${\hat\rho}^{a,p}$ to Eq.~(1). ${\hat\sigma}^{a,p}$ were obtained by numerically inverting ${\hat\rho}^{a,p}$ tensor. Comparing these calculated results with the measured drag resistivity shown in Fig.~2(a-b), we find theory provide reasonable match to experiment by capturing key features of the sign and magnitude of the observed drag. To be specific, for $\rho_{xx}^{drag}$, the calculation successfully captured that the sign of drag is governed by carrier types and does not change cross LLs for graphene EDL specifically. For $\rho_{xy}^{drag}$, the complicated changes of Hall drag signs are also revealed by the calculation. We note that while the calculated drag exhibits excellent agreement with the data in the compressible regime, the agreement between experiment and calculation is worse in the incompressible strips, especially for $\rho_{xy}^{drag}$. Specifically, the measured drag signals vanish as expected while the calculated one do not. This deviation is due to the imperfect measurement geometry for $\rho^{top}_{xx}$ and  $\rho^{bot}_{xy}$ due to less ideal device geometry, which the plateau of QH to be not fully developed (as can be seen in fig.2f, g). This non-perfect quantization results in finite $\frac{d\hat{\sigma}}{dn}$ which lead to non-zero calculated drag.

While comparing the absolute magnitude of experimental drags to theoretical expectation is not possible due to the undetermined prefactor in Eq.~(1), we can still make a relative comparison of the magnitude of different components of the drag resistivity tensor. Fig.~2(i) shows an example of such a comparison along the equal-density line ($n_{T}= n_{B}$). Note that the prefactor in Eq.~(1) could be a function of density, temperature and field. We multiplied a common factor to Eq.~(1) to make the calculated results comparable to experimental ${\hat\rho}^{drag}$. Plotting the magneto-drag and Hall drag in the same scale, we found that the relative magnitude between measured $\rho^{drag}_{xx}$ and $\rho^{drag}_{xy}$ (solid curves) match well with the calculation (dashed curves), proving that Eq.~(1) holds quantitatively. For best matching, we also note that we multiplied different common factors for different LLs whose ratio is $\sim$1.3 for N=0 to N=1 Landau level (separated around $V_{BG}\approx$10~V, indicating that the prefactor in Eq.~(1) can be the LL filling fraction dependent but with a weak density dependence within a LL.


 \begin{figure}

 \includegraphics[scale=0.46]{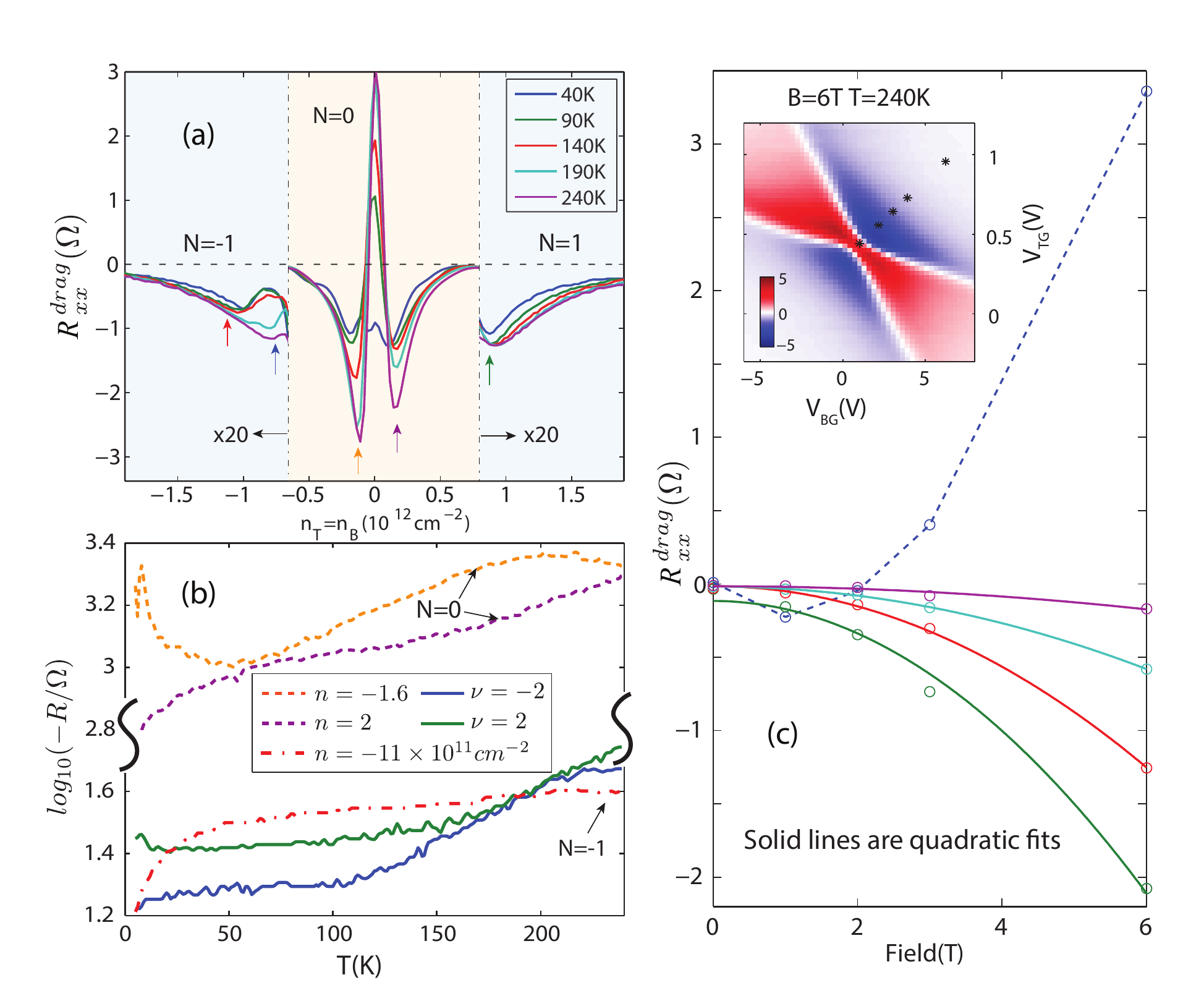}
 \caption{(a)Magneto-drag along equal density line ($n_{B}=n_{T}$) as a function of densities at field of 13~T and different temperatures. The blue shaded region marks $N=\pm1$ LL and the yellow shaded marks $N=0$ LL. The drag signals are multiplied by a factor of 20 in the blue shaded regions for clarity. (b) drag as a function of temperature at different density points. Solid lines represent the Landau gaps $\nu=\pm2$. Dashed line represent partially filled LLs: $n=-1.6, 2\times 10^{11}$cm$^{-2}$ corresponding to $N=0$ LL; $n=-11\times 10^{11}$cm$^{-2}$ correspond to $N=-1$ LL. The density of each line in (b) is marked out in (a) by arrows with corresponding colors. (c)Magneto-drag as a function of field at temperature of 240~K, at certain density points along equal density line (shown as * in the insert). Circles are experimental data and solid curve are quadratic fit of the data.}{\label{F3}}
 \end{figure}

Finally, we discuss the temperature and magnetic field dependence of drag signals. Unlike the zero magnetic field drag, which is found to be proportional to $T^2$ owing to the increasing scattering phase space in the Fermi liquid~\cite{Gorbachev2012}, $\rho_{xx}^{drag}$ measured in the high magnetic field regime exhibits a relatively weak temperature dependence. Fig.~3(a) and (b) show the temperature and density dependent $\rho_{xx}^{drag}$. We note that even for $N=0$ LLs (orange shaded region in Fig.~3(a)), where we observed the most significant temperature dependence, the drag signals increase only by a factor of $\sim$2 as temperature changes from 40 K to 240 K. In particular, when both layers are on $N=\pm1$ Landau level (red dashed line in Fig.~3 (b), there is almost no temperature dependence above $\sim$40K. The observed temperature insensitive drag effect is presumably due to the fact that the thermal energy is much larger than the individual LL spreads, but much smaller than the LL spacing (cyclotron gap). When temperature is much smaller than the cyclotron gap, only one LL is partially occupied while the LLs above or below are completed empty or full. And if temperature is much larger than LL spread, the entire partially filled LL is always accessible for Coulomb scatter. In this temperature regime, temperature no longer controls the scattering phase space, so drag no longer depends on temperature. Interestingly, $\rho_{xx}^{drag}$ exhibits a strong magnetic field dependence. Fig.3c shows $\rho_{xx}^{drag}$ as a function of magnetic field at T=240K, where a $B^2$ dependence is observed across different densities. One possible explanation of the strong field dependence is that the scattering phase space is enlarged by the increase of the LL degeneracy at higher fields.


In conclusion, we measured magneto- and Hall-drag in graphene double layer in the quantum Hall regime in the presence of strong thermal fluctuations. We observed strong drag signals, which vanish when either layer is in the incompressible quantum Hall state.  The noted magneto- and Hall-drag are well described by the variation of magneto transport tensors in the drive and drag layer, confirming the theory proposed by OSS. This also indicates frictional momentum drag is the dominant mechanism for Coulomb drag in graphene EDL under strong magnetic field.

We thank Felix von Oppen, Ady Stern and Bertrand Halperin for helpful discussion. The major experimental work is supported by DOE (DE-SC0012260). P.K. acknowledges partial support from the Gordon and Betty Moore Foundation's EPiQS Initiative through Grant GBMF4543. K.W. and T.T. acknowledge support from the Elemental Strategy Initiative conducted by the MEXT, Japan. T.T. acknowledges support from a Grant-in-Aid for Scientic Research on Grant262480621 and on Innovative Areas Nano Informatics (Grant 25106006) from JSPS.

\bibliography{Coulombdrag}

\end{document}